# Screening and evaluation of potential clinically significant HIV drug combinations against SARS-CoV-2 virus

Draško Tomić[1*], Karolj Skala[1], Attila Marcel Szasz[2], Melinda Rezeli[3], Vesna Bačić Vrca[4], Boris Pirkić[5], Jozsef Petrik[6], Vladimir Janđel[7], Marija Milković Periša[8], Branka Medved Rogina[9], Josip Mesarić[10], Davor Davidović[1] and Tomislav Lipić[10]

[1]Center for Informatics and Computing, Ruđer Bošković Institute, Zagreb, Croatia; [2]Department of Bioinformatics and Cancer Center, Semmelweis University Budapest, Hungary; [3]Department of Biomedical Engineering, Lund University, Lund, Sweden; [4]Central Hospital Pharmacy, Clinical Hospital Center Dubrava, Zagreb, Croatia; [5]Surgery, Orthopedics and Ophthalmology Clinic, Faculty of Veterinary Medicine, University of Zagreb, Zagreb, Croatia; [6]Department of Medical Biochemistry and Hematology, Faculty of Pharmacy and Biochemistry, University of Zagreb, Zagreb, Croatia; [7]Deprtment of Obstetricy, Clinical Hospital Centre Zagreb, [8]Institute of Pathology, School of Medicine University of Zagreb, Zagreb, Croatia; [9]Laboratory for Information and Signal Processing, Division of Electronics, Ruđer Bosković Institute, Zagreb, Croatia; [10]Faculty of Electrical Engineering and Computing, University of Zagreb, Zagreb, Croatia; [11]Machine Learning Lab and Knowledge Representations, Division of Electronics, Ruđer Bošković Institute, Zagreb, Croatia.

**Abstract:** In this study, we investigated the inhibition of SARS-CoV-2 spike glycoprotein with HIV drugs and their combinations. This glycoprotein is essential for the reproduction of the SARS-COV-2 virus, so its inhibition opens new avenues for the treatment of patients with COVID-19 disease. In doing so, we used the VINI in silico model of cancer, whose high accuracy in finding effective drugs and their combinations was confirmed in vitro by comparison with existing results from NCI-60 bases, and in vivo by comparison with existing clinical trial results. In the first step, the VINI model calculated the inhibition efficiency of SARS-CoV-2 spike glycoprotein with 44 FDA-approved antiviral drugs. Of these drugs, HIV drugs have been shown to be effective, while others mainly have shown weak or no efficiency. Subsequently, the VINI model calculated the inhibition efficiency of all possible double and triple HIV drug combinations, and among them identified ten with the highest inhibition efficiency. These ten combinations were analyzed by Medscape drug-drug interaction software and LexiComp Drug Interactions. All combinations except the combination of cobicistat_abacavir_rilpivirine appear to have serious interactions (risk rating category D) when dosage adjustments/reductions are required for possible toxicity. Finally, the VINI model compared the inhibition efficiency of cobicistat_abacivir_rilpivirine combination with cocktails and individual drugs already used or planned to be tested against SARS-CoV-2. Combination cobicistat_abacivir_rilpivirine demonstrated the highest inhibition of SARS-CoV-2 spike glycoprotein over others. Thus, this combination seems to be a promising candidate for the further in vitro testing and clinical trials.

# 1. INTRODUCTION

It is still unclear where the SARS-CoV-2 virus came from, and we have no scientific confirmation that the virus has passed from an animal species to a human one. However, we do have confirmation of the transmission of SARS-CoV-2 virus from humans to dogs and cats. Specifically, WHO (World Health Organization) data describes two dogs in Hong Kong and one cat in Belgium, which were virus-contaminated from humans, but this was not COVID-19 infection in these animals. Shortly after the first confirmed cases of infection with SARS-CoV-2 coronavirus [1], COVID-19 spread rapidly to almost all continents of the world, with the disease becoming a pandemic in a very short time [2]. The number of patients confirmed to be infected with SARS-CoV-2 is increasing rapidly, and at the time of submission of this work, according to WHO [3], it was more than 2,500,000 cases. The end of this pandemic cannot yet be foreseen, and to make matters even worse, the estimated mortality caused by this virus is from about 1% to as high as 12% in the epicentre of the epidemic [4].

A host of high frequency mutations have resulted in at least five differentiated SARS-CoV-2 strains to date [5], therefore finding an effective vaccine against this virus is not to be expected soon. Consequently, efforts by the world scientific community and the pharma industry need to be focused on finding effective therapies as quickly as possible, in order to reduce the number of deaths. Finding new antiviral drugs is generally a costly and time-consuming process, and often limited with our understanding of biology [6]. Therefore, it is justified to consider the use of existing drugs to cure new diseases [7], as these drugs already have well-established recommended doses and regimens, known side effects, and ways of preventing or mitigating such effects. Equally important, the optimal methods of synthesizing existing drugs are known, so in the case of increased demand for a certain drug, it is easier, faster, and cheaper to expand existing production capacities than to design and build new ones [8]. Repurposing existing drugs to treat diseases can provide additional important benefits, and this is the possibility to administer multiple drugs at the same time. For example, by using combinations of drugs whose efficacy against the specific mechanism of pathogen is greater than the efficacy of the individual drugs, it is possible to achieve greater therapeutic efficacy. Another approach is to use the combination of several drugs with each drug acting on the different mechanism of pathogen [9]. In addition, the combined use of multiple drugs may reduce the potential for the pathogen to develop resistance [10]. This approach is now standard in the treatment of many serious diseases, including the treatment of cancer [11], bacterial infections [12], and infections with HIV [13].

The first attempts at treating COVID-19 disease with existing drugs have already been made. Thus, Kaletra (lopinavir/ritonavir combination) alone or in combinations with α-interferon, reverse transcriptase inhibitors (emtricitabine/tenofovir alafenamide fumarate), oseltamivir, and guanosine analog and reverse transcriptase inhibitor ribavirin, is being tested on SARS-CoV-2 patients. Trials with ritonavir plus ASC09, umifenovir, and remdesivir are also planned [14]. Antiviral drug favipiravir, chloroquine, and nucleotide analog remdesivir are under investigation [15].

Given the research to date and clinical findings, we decided to systematically investigate the possibility of administering pre-existing antiviral medicines and their combinations in the treatment of COVID-19. As a tool for our investigation, we chose the VINI in silico model of cancer [16]. The VINI model performs virtual drug screening [17] on KEGG diseases metabolic pathways [18]. The high accuracy of the VINI model in calculating the efficacy of cancer drugs and their combinations has already been confirmed by comparison with in vitro NCI-60 data and clinical trials [16] [19]. Furthermore, its modular architecture enables its easy extensibility and virtual drug screening of nearly all diseases described by KEGG metabolic pathways. However, since there is no metabolic pathway for COVID-19 disease until now established, we have taken the advantage of the VINI model's ability to perform virtual drug screening also on only one specific protein.

Like other coronaviruses, SARS-CoV-2 has four structural proteins, Figure 1, known as the S (spike), E (envelope), M (membrane), and N (nucleocapsid) proteins; the N protein holds the RNA genome, and the S, E, and M proteins together create the viral envelope [20].

Although all four of these structural proteins are potential drug targets, we restricted our investigation to spike protein, which is essentially a glycoprotein that allows the binding of SARS-CoV-2 to ACE2 (an angiotensin converting enzyme 2) and the transfer of viral RNA material to the host cell [21]. The three-dimensional structures of spike glycoprotein were deposited by the same author's team at www.rcsb.org [22] on 11/03/2013. In our study, we used one of these structures, designated 6VXX [23], as a target for virtual drug screening.

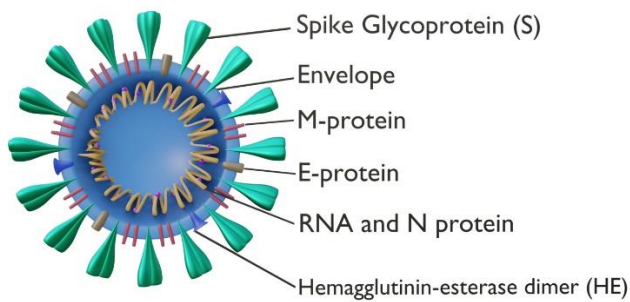

Figure 1. Structural image of SARS-CoV-2. Glycoprotein Spike on the outer edge of virus particles give coronaviruses their name, crown like.

*Address correspondence to this author at the Center of Informatics and Computing, Rudjer Boskovic Institute, 10000 Zagreb, Croatia; Tel ++385 - 91-577-8380; e-mails: drasko@irb.hr

## 2. METHODS AND RESULTS

The task of virtual drug screening is to find drugs or chemical compounds (ligands) with high free energy of binding (in further text abbreviated ΔG) with target molecules (receptors), thus inhibiting them. The VINI model uses Autodock Vina software [24] to calculate ΔG [25] between receptors and ligands. By definition, ΔG has always negative values expressed in kcal/mol units, or zero value in case there is no binding between receptor and ligand. The lower these negative values, the greater the inhibition of the receptor by ligand is. For example, the binding energy of the compound A with ΔG = -15 kcal/mol is considered to have higher inhibition than the compound B with the binding energy ΔG = -10 kcal/mol. Although receptors can be any type of chemical compounds, most common receptors in biological processes are proteins. Unlike receptors, drug candidates are mostly small molecules composed of several to several dozen atoms.

The quest for a vaccine against the novel SARS-CoV-2 is recognized as an urgent problem [26], however the development of an effective SARS-CoV-2 vaccine is not expected soon. For this, at this stage of the rapidly growing COVID-19 pandemic one of the most important tasks is to find effective drugs against SARS-CoV-2 as soon as possible. Various approaches were suggested, for example to use available angiotensin receptor 1 (AT1R) blockers, such as losartan, and block angiotensin-converting enzymes (ACE2) on human cells, thus preventing SARS-CoV-2 from attaching to ACE2 with its spike glycoprotein [27].

Our approach is opposite: to block SARS-CoV-2 spike glycoprotein directly with the combination of available antiviral drugs expressing highest inhibition to it, thus preventing the virus from attaching to the human cell. Therefore, we decided to perform a virtual screening of the efficacy of the existing and approved antiviral drugs and their combinations on the SARS-CoV-2 spike glycoprotein.

On KEGG portal we identified 52 FDA-approved antiviral drugs, 44 of them small molecules with an experimentally determined three-dimensional (3D) structure, 7 interferon drugs with FASTA sequence, and sinecatechin, a specific water extract of green tea leaves from Camellia sinensis. Furthermore, regardless of the fact that virtual drug screening with reduced accuracy is possible to perform on proteins without structure by predicting their 3D structures from FASTA sequences [28], we omitted interferon drugs from this study due to the extremely high demand of molecular dynamic tools [29] needed to perform in silico protein-protein docking and calculation of their binding free energies [30].

Of these small-molecule anti-viral drugs, we found 22 drugs indicated for human immunodeficiency virus (HIV), 6 for influenza and herpes simplex virus (HSV), 4 drugs for hepatitis B virus and cytomegalovirus, and 2 for respiratory syncytial virus (RSV). Additionally, we excluded from the further analysis antiviral drugs whose binding energy to SARS-CoV-2 is weak. The VINI model first calculated ΔG of 44 small-molecule anti-viral drugs with SARS-CoV-2 spike glycoprotein. As for all other calculations of ΔG in this study, the VINI model used Autodock Vina software, performed docking and binding free energy calculation between SARS-CoV-2 spike glycoprotein structure file with RCSB pdb id 6VXX and each small-molecule drug 10 times, and then computed average of ΔG expressed in kcal/mol units. For the purposes of this and all subsequent simulations, the VINI model pulled a 6VXX structure file from the RCSB portal. For antiviral drugs, the VINI model used 3D structures from the Drugbank [31]. The results of these calculations are presented in Figure 2 (see the details of experimental results in Table I in the section Supplementary Material).

A higher negative value indicates a higher energy of binding. Values below 7.0 kcal/mol represent weak binding energies. The highest ΔG values were shown by two HIV drugs atazanavir and saquinavir, and the weakest by foscarnet, a cytomegalovirus drug. Abbreviations used: CMV (cytomegalovirus), IFV (influenza virus), HSV (herpes simplex virus), HIV (human Immunodeficiency virus), HBV (hepatitis B virus), and RSV (respiratory syncytial virus).

In the case of multi-drug cocktail therapy, the final binding energy depends on the order of drug binding to the protein. For example, the binding energy ΔG of protein P first with drug A and then drug B differs from the binding energy of protein P with drug B and then with drug A. Symbolically, this can be written as:

$$\Delta G(P, A, B) \neq \Delta G(P, B, A) \qquad (1)$$

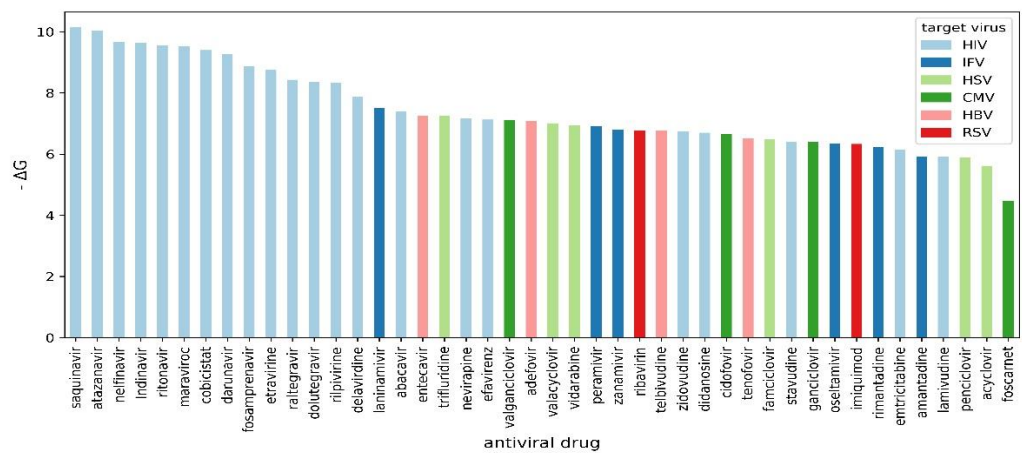

Figure 2 Presented are calculated -ΔG values between the SARS-CoV-2 spike glycoprotein and the small-molecule antiviral drugs, expressed in kcal/mol units.

Furthermore, in the present study, and due to the possible adverse drug interactions, we limited our search for cocktail therapies with up to three individual drugs. The total number of single, double and triple combinations with N drugs equals can be expressed as:

$$M = N + N^2 + N^3 \quad (2)$$

From (2) and because Autodock Vina simulation between SARS-CoV-2 spike glycoprotein and each drug was performed 10 times, it follows that for all single, double and triple combinations with N = 44 drugs the total number of required simulations will be 10*M, which equals 871,640. We estimated that such a calculation on the computing infrastructure available for this study may take more than two weeks. However, such a long calculation time is unacceptable given the situation caused by the SARS-CoV-2 virus, which requires the rapid development of new drugs.

Because of this, the fact that HIV drugs and their combinations are most frequently represented in previous attempts to treat COVID-19 disease, and of the results given in Figure 2 showing that most HIV drugs have a binding energy to the SARS-CoV-2 glycoprotein higher than other antiviral drugs, we further limited the scope of virtual drug screenings to HIV drugs only.

Consequently, the VINI model calculated binding free energies for all double and triple combinations of HIV drugs, and isolated ten combinations with the highest inhibition of SARS-CoV-2 spike glycoprotein. The results of these calculations are shown in **Error! Reference source not found.** (see the details of experimental results in Table II in the section Supplementary Material).

All calculated values refer to -ΔG between SARS-CoV-2 glycoprotein and drugs, and are given in kcal/mol units. Drug-drug interactions and possible side effects of single drugs were examined using Medscape [32] and Lexicomp [33] drug interaction checkers. Combinations failing into category D (it is not recommended to administer these combinations without reducing the dosage of the individual drugs) are indicated in red. Combination cobicistat_abacavir_rilpivirine is highlighted green, as both Medscape and Lexicomp found no serious interaction effects. That means, no dosage reduction of the individual drugs is required (category C). The efficiency ($E$) of a particular combination of drugs or a single drug is defined as the absolute sum of the lowest (min) and highest (max) total binding energy:

$$E = |\min(\Delta G) + \max(\Delta G)| \qquad (3)$$

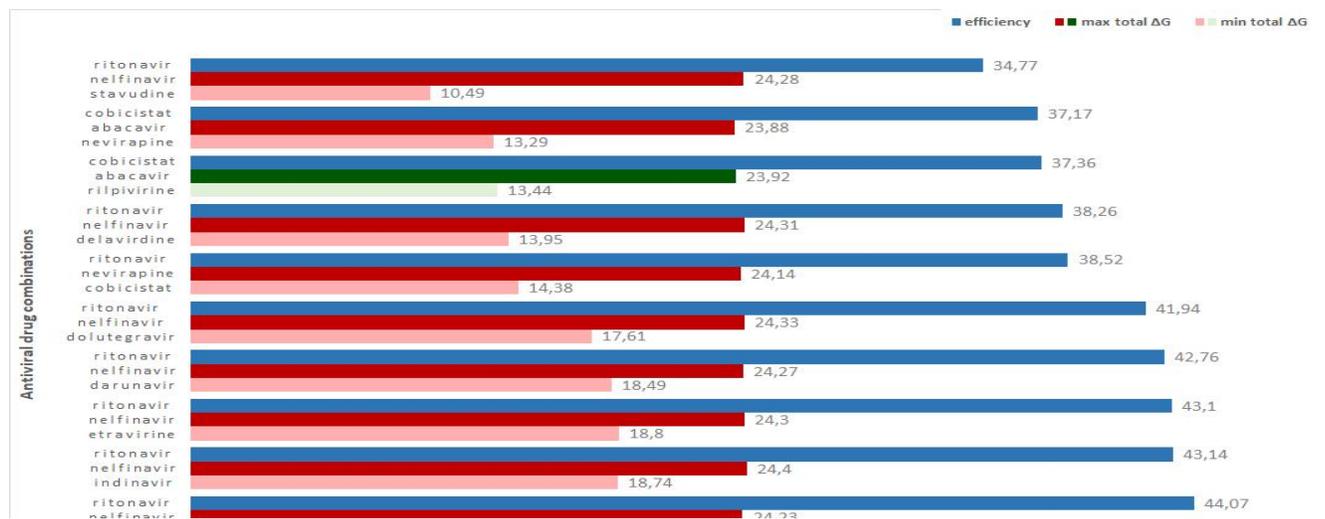

**Figure 3**. 10 combinations found by the VINI model to be the most effective in inhibiting SARS-CoV-2 spike glycoprotein.

In the case of single drugs, the min and max binding energies are the same, and therefore their efficiency is absolute value of $2 * \Delta G$.

In order to compare the relative efficacy of cobicistat_abacavir_rilpivirine cocktail in inhibiting SARS-CoV-2 spike glycoprotein against cocktails and single drugs already used to treat COVID-19 or under the investigation, the VINI model calculated ΔG for the following combinations and compounds: Kaletra alone, Kaletra in combinations with oseltamivir, remdesivir and ribavirin, and hydroxychloroquine, remdesivir, favipiravir and umifenovir alone. The results of this comparison are presented in Figure 4 (see the details of experimental results in Table III in the section Supplementary Material).

Drug-drug interactions and possible side effects of single drugs were examined using Medscape and Lexicomp software. The cocktails and single drugs that fall into category C are indicated in green. This category does not require a reduction in the doses of the individual drugs in combination. The combination of Kaletra with remdesivir and remdesivir alone are highlighted in red because remdesivir is investigational antiviral compound, is not yet licensed or approved anywhere globally, and has not been demonstrated to be safe or effective for any use.

The predicted efficiency, computed as in (3), of cobicistat_abacavir_rilpivine combination is higher than the efficacy of other cocktail therapies currently used as of our best knowledge against SARS-CoV-2. The gain of cobicistat_abacavir_rilpivine cocktail in inhibiting SARS-CoV-2 spike glycoprotein over Kaletra is 13,62%, over Kaletra with oseltamivir cocktail 7,25%, and over Kaletra with ribavirin cocktail 6,98%. The gain over single drugs used or planned to be used is much higher, and is 69,27% over favipiravir, and 61,67% over umifenovir. The gain over hydroxychloroquine is not relevant because this drug works on a completely different basis, by suppressing the immune system overreaction. Only the predicted efficacy of Kaletra with remdesivir cocktail is higher than of cobicistat_abacavir_rilpivirine combination, and equals 4,83%. However, remdesivir is not an approved drug, but a compound still under investigation in recently approved clinical trials.

From the results, one can also discern a slight fluctuation of ΔG results for the same receptor-ligand pair in different experiments. Thus, regardless of the fact that ΔG was calculated ten times for each receptor-ligand pair, in the first experiment ΔG for indinavir was -9.63 kcal / mol (Table 1), and in the second experiment ΔG was -9.50 kcal / mol (Table 2). Such fluctuations are expected and are related to the stochastic nature of Autodock Vina, which each time initiates simulation of the same receptor-ligand pair from different, randomly selected initial coordinates.

## 3. DISCUSSION AND CONCLUSION

The SARS-CoV-2 spike glycoprotein is a key factor in the binding of the virus to angiotensin-converting enzyme (ACE2), allowing the virus to transmit its RNA material to the host cell, and hence its further reproduction. Effective inhibition of the SARS-CoV-2 spike glycoprotein can slow or even completely inhibit the reproduction of the SARS-CoV-2 virus, thus increasing the chances of the host immune system to fight against the virus. The results of our study provide scientific confirmation of the high efficacy of some combinations of approved HIV drugs, and compounds still under investigation, in the inhibition of SARS-CoV-2 spike glycoprotein. Further application of these results may pave the way for new, even more effective therapies than those currently used in the treatment of COVID-19.

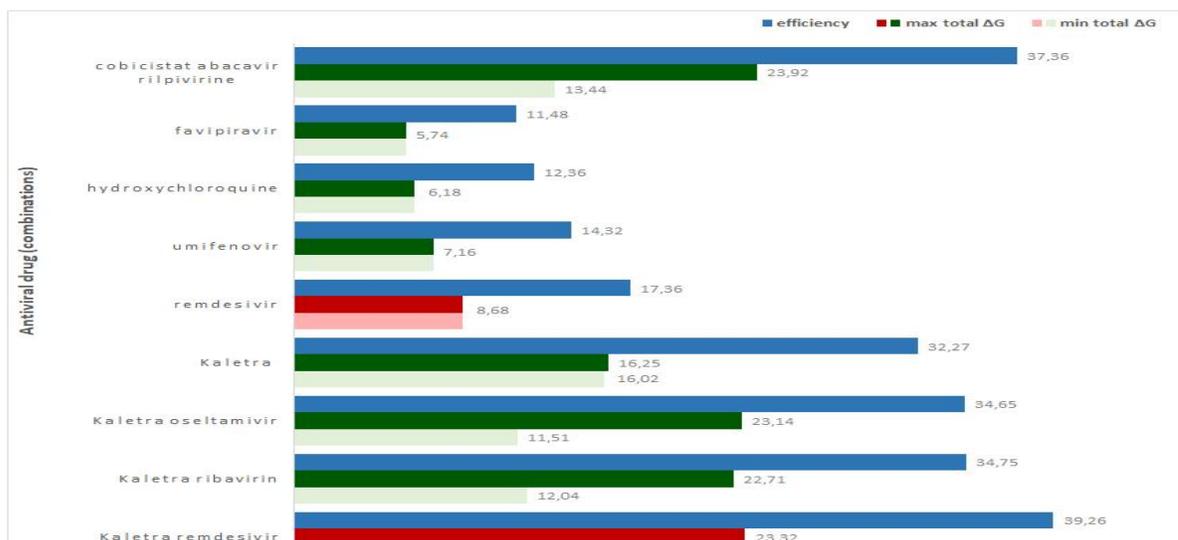

Figure 4. An overview of the efficacy of cobicistat_abacavir_rilpivine combination of HIV drugs with respect to the already used or planned combinations of drugs or individual drugs.

Among the ten most effective HIV drug combinations predicted by the VINI model to be the most effective in inhibiting SARS-CoV-2 spike glycoprotein, the cobicistat_abacavir_rilpivirine combination stands out. Although ranked 8[th] regarding its efficiency, it is, due to its relatively low toxicity and acceptable drug-drug interactions, suitable for use at standard doses of individual drugs without the need for their adjustment.

The predicted efficacy of this combination in the inhibition of SARS-CoV-2 spike glycoprotein is greater than the efficacy of the drug combinations and the individual drugs, which, to the best of our knowledge, have been tried, or are planned to be tried in the treatment of COVID-19. These are Kaletra (lopinavir_ritonavir combination), Kaletra in combination with oseltamivir, Kaletra in combination with ribavirin, and hydroxychloroquine, favipiravir and umifenovir as single drugs.

The efficiency of the cobicistat_abakavir_rilpivirin combination predicted by the VINI model is slightly lower than the efficacy of the Kaletra_remdesivir combination. However, remdesivir is a compound under investigation, and its possible side effects and interactions with lopinavir and ritonavir are still unknown.

From the results obtained, it can also be concluded that the order of action of individual drugs in some combination on the SARS-CoV-2 spike glycoprotein does matter. Thus, the highest ΔG of cobicistat_abacavir_rilpivirin combination on this glycoprotein is achieved when it first binds cobicistat, then abacavir, and at the end rilpivirine. This indicates that a 3 times daily regimen in which cobicistat is given first, then abacavir, and at the end rilpivirine, could further increase the effectiveness of this combination therapy. A detailed pharmacokinetic analysis of the effects of individual drugs will be required to evaluate the benefits such a regimen may possibly bring, and this goes beyond the scope of this study.

Therefore, a clinical trial of the combination of cobicistat_abacavir_rilpivirin with a limited number of COVID-19 patients who are in moderate severe and severe condition is warranted.

**CONFLICT OF INTEREST**

The authors declare that the research was conducted in the absence of any commercial or financial relationships that could be construed as a potential conflict of interest.

**AUTHOR'S CONTRIBUTION**

Tomic D: The development of the VINI in silico model of cancer and its extension towards COVID-19 disease. Screening the antiviral drugs and their combinations for their inhibition against SARS-CoV-2 spike glycoprotein. Prepared first draft version of the manuscript; Karolj S: Organizing the paper and synchronizing the co-authors. The review and proofreading of the paper. Adding visualization solutions and new drug combinations for virtual screening; Szasz A.M: Discussed the results and contributed to the writing; Rezeli M: Discussed the results and contributed to the writing; Bacic Vrca V: Screening and analysis of potential clinically significant drug - drug interactions; Pirkic B: Analysis of possible routes of SARS-CoV-2 virus transmission from humans to animals and vice versa. Discussed the results and contributed to the writing. Petrik J: Produced the original picture of SARS-CoV-2. Screening and analysis of potential clinically significant drug - drug interactions; Milkovic Perisa M: Discussed the results and contributed to the writing; Medved Rogina B: Discussed the results and contributed to the writing; Mesaric J: Preparation of the graphics; Davidovic D: Optimization of the VINI model code. Discussed the results and contributed to the writing; Lipic T: Exploratory analysis and visualizations. Discussed the results and contributed to the writing.

**ACKNOWLEDGMENT**

Simulations in this research were performed on the supercomputer Bura at the University of Rijeka, Croatia, which was procured under the project "Development of research infrastructure at the University campus in Rijeka", co-funded by the European Regional Development Fund (ERDF).

**DATA AVAILABILITY STATEMENT**

The datasets generated for this study and the VINI in silico model of cancer can be found in the Full-text Institutional Repository of the Ruđer Bošković Institute, http://fulir.irb.hr/5420/.